# Cooperation Networks: Endogeneity & Complexity[*]

## Simon Angus[†]

### July 2006

Manuscript Only (Under Review)

### Abstract

Insights from the Complex Systems literature are employed to develop a computational model of truly endogenous strategic network formation. Artificial Adaptive Agents, implemented as Finite State Automata (FSA), play a modified two-player IPD game with an option to further develop the interaction space as part of their strategy. Several insights result from this minor modification: first, I find that network formation is a necessary condition for cooperation to be sustainable but that both the frequency of interaction and the degree to which edge formation impacts agent mixing are both necessary conditions for cooperative networks. Second, within the FSA-modified IPD frame-work, a rich ecology of agents and network topologies is observed and described. Third, the system dynamics are investigated and reveal that initially simple dynamics with small interaction length between agents gives way to complex, a-periodic dynamics with self-organized critical properties when interaction lengths are increased by a single step.

---

[*]This work was initiated at the Santa Fe Institute Graduate Workshop on Computational Social Sciences and Complexity, Santa Fe, New Mexico. My sincere thanks go to John Miller and Scott Page and the other participants for their input, advice and tuition over this stimulating time; in particular, to John, for his warm guidance on this project. All errors in this work, whether factual, computational, or other, are solely the responsibility of the author.

[†]*Post: School of Economics, John Goodsell Building, the University of New South Wales, 2052 NSW, Australia; E-mail (preferred):* s.angus@unsw.edu.au; *Phone +61 2 9385 3334; Fax +61 2 9313 6337.*

## 1 INTRODUCTION

The strategic literature has seen a long-standing interest in the nature of cooperation, with many contributions considering the simple but insightful two-player Prisoner's Dilemma (PD) game,

$$
\begin{array}{c}
\textbf{II} \\
\begin{array}{c|cc}
 & C & D \\
\hline
C & (a,a) & (b,c) \\
D & (c,b) & (d,d)
\end{array}
\end{array}
\qquad (1)
$$

where $c > a > d > b$ and $a > (b+c)/2$.[1] Traditionally,[2] such games were analysed under an uniform interaction specification such that agents met equiprobably to play a single (or repeated) two-player game such as in (1).[3] In graph-theoretic terms, such matching can be thought of as a choice of one edge out of $n(n-1)/2$ edges in a complete graph of size $n$ (Fig. 1). More recently however, authors have relaxed this condition, and have analysed strategic games of cooperation and coordination under both non-uniform interaction and non-uniform learning environments. Here, some form of topological imposition, other than the complete graph, such as uni-dimensional play on a line, circle, or higher-dimensional interaction on a regular graph (e.g. a torus) is usually applied, with more recent contributions allowing for richer (statistical) graph environments such as so-called 'small-world' graphs.[4] The topological significance of the interacting space has been stressed by these authors as it appears to influence the degree to which cooperation can be sustained. For instance, Anderlini and Ianni (1996) and Anderlini and Ianni (1997) find that different actions of a pure coordination game survive in the long-run at different locations on the interaction space; whilst Kirchamp (2000) computationally studies interacting agents on a torus playing the PD and coordination games, with cooperation and non-risk-dominant coordination outcomes observed respectively. Consequently, and reflecting a burgeoning interest in networks of all kinds, much attention has been paid to the study of realistic social networks (Barabasi et al., 2002; Baum et al., 2003), with statistical network characterisation (Watts and Strogatz, 1998; White et al., 2004) and clique analysis similarly receiving interest (Girvan and Newman, 2001; Tyler et al., 2003).

However, apart from a few exceptions[5], authors have not allowed the interaction environment itself to vary, either (for example) due to some exogenous schedule or as a result of processes endogenous to the model. Clearly, such modelling features comprise an highly desirable step towards treating realistic economic and social networks.[6] In this vein, two directions of causality are apparent, first with respect to how a changing interaction environment might affect strategic outcomes for agents (the 'topological' effect); and second, how agents *through their own strategic actions* might impact on the very interaction space itself (the 'agency' effect).

---

[1] As in Axelrod and Hamilton (1981).

[2] See for example, within an evolutionary game theoretic framework, Kandori (1992).

[3] In the following discussion, we shall refer to this case as 'uniform interaction'.

[4] For example, Elgazzar (2002); Kirchamp (2000); Masuda and Aihara (2003); Stocker et al. (2002).

[5] See for example Bala and Goyal (2000); Dutta et al. (1998); Ely (2002); Jackson and Watts (2002); Mitchell (1999); Slikker and van den Nouweland (2000).

Such a system of interacting agents, with heterogeneity in agent behaviour *and* interaction profiles fits well into the so-called 'science of complexity'.[7] This approach seeks to identify and study systems whose components interact in some non-uniform, (and usually) non-linear manner. In particular, due to the inherent unpredictability of such non-equilibrium systems, agent-based computational modeling techniques provide an extremely useful method of enquiry especially where non-rational learning behavior is also suspected (Arthur, 1994; Holland and Miller, 1991).[8]

The current paper reflects such an approach. Specifically, both constraints concerning agent rationality and rigid agent interactions are relaxed within a fundamentally agent-based modelling framework. Moreover, in contrast to one related approach in the literature (see below), agents are given *strategic* abilities to change the interaction space *themselves* (i.e. to change interaction probabilities) during pair-wise game-play. It is in this sense that a 'network' arises in the model, and hence, such a network is said to be a truly *endogenous* feature of the modelling framework; a feature which to my knowledge has not been previously handled with boundedly rational agents.

The key insights of the present work can be summarised as follows: first, an analytic analysis without network formation reveals that the modification to the standard iterated PD (IPD) framework introduced below does not change the canonical behaviour of the system; second, that when network formation is afforded, stable cooperation networks are observed, but only if both a type-selection and enhanced 'activity' benefit of the network is present; third, that the extended system under certain interaction lengths is inherently self-defeating, with both cooperation and defection networks transiently observed in a long-run specification; and fourth, that the network formation process displays self-organized criticality and thus appears to drive the complex dynamics observed in the long-run.

The rest of the paper is organised as follows: first, a discussion of related literature is presented; second, the model is introduced, paying particular attention to the modelling of agents and incorporation of network form-ing behaviour; third, analysis is performed analytically on the basic (non-network forming) model before extension to incorporate network formation is performed on both a short- and long-run time horizon; and finally, some concluding observations and a discussion of possible extensions is made.

---

[6]Consider, for example, the *guanxi* business network in China documented in (Standifird and Marshall, 2000) and (Fan, 2002).

[7]Non-technical introductions can be found in (for example) Lewin (1993) or Coveney and Highfield (1996), or a survey of economic applications is found in Tesfatsion (2003).

[8]See also applications of artificial adaptive agents to organizational problems in Choi (2002); Marsili et al. (2004); Stocker et al. (2002).

## 2  RELATED LITERATURE

The current specification, where a non-uniform interaction structure is allowed, is related in intention to the preferential partner selection (*choice*) and optional rejection of an offer to interact (*refusal*) literature (or IPD/CR when the game is the IPD). Here, the emphasis is on how the added mechanism of choice and refusal affects the emergence of cooperation in IPD games. Such a mechanism is seen as more realistic, from both a biological, and social perspective[9]. For example, Ashlock and co-workers 1996 construct a computational model (see below) to consider the effects on cooperative behaviour with varying levels of preferential selection, finding that most ecologies converge to full cooperative behaviour but that 'wallflower' ecologies are possible if intolerance to defection is high, or costs to social exclusion is low.[10] Such findings are supported to some extent by the experimental work of Hauk and Nagel (2001) who find that cooperative behaviour increases over time under unilateral choice of partners (opponent must accept to play).

Similarly, authors have considered cooperation (or corruption) arising in informal networks. In these studies, a 'network' is used to describe a (proper) subset of agents in the population who are then distinguished from the majority in some way. Taylor's 'old-boy network' model (Taylor, 2000) studies networked agents to be those of a certain type – the qualified/competent type. Membership of this network is conferred upon the individual after 'showing their colours' in an interaction. The mixing of agents is population-wide, and therefore, in this model, the 'network', although giving important *type* information for future transactions, plays no more part in the interaction space, nor does the actual topology of the network matter.[11] Since there is no network exit criterion, nor behavioural dynamic, Taylor finds that networks are rarely socially optimal (as opposed to anonymous transactions) since a bleeding of the 'good' types from the general population ensues (compare (Kali, 1999)).

However, these approaches suffer from the constraints imposed by the analytic framework, thus only allowing one (informal) connected component to form with such formation not endogenised; authors assume that where networks are sustainable they will form.

---

[9] See the introduction to Smucker et al. (1994) on such observations.

[10] See also, Tesfatsion's work on trading games with endogenous partner selection (Tesfatsion, 1997).

[11] As is perhaps clear, this is not a network in the sense of a formal graph with an edge set, but can be thought of as a disjoint graph with the 'network' comprising a complete connected component.

Perhaps the closest work to the current paper, and bridging the IPD/CR – endogenous network literature, is a second paper by Smucker et al. (1994). Here, a similar computational model to that previously mentioned is used, but in addition to considering the strategic implications of various levels of choice and refusal, they also perform some characterisation of the evolving network of interactions. In the SSA model, agents are modeled as 16 state Moore machines[12] who are programmed to play the IPD. However, and significantly for the present study, the 'network' in the SSA model is defined by a simple global rule – if the number of interactions between two players is (statistically) significantly larger than the mean interaction count for the whole population, then an edge is assigned between these players. Thus, for SSA, the 'network' is more a record of 'acceptable payoff outcomes' rather than a functional entity which shapes future interactions. This is an acknowledged limitation of the work.

The present work aims to address many of the mentioned shortcomings of the literature. First, by implementing network formation as a strategic and therefore inherently endogenous process; second, and following on from the first, by allowing for multiple networks to form simultaneously (rather than one connected component only); and third, by implementing agent strategies as finite state automata, both bounded rationality and learning are incorporated.

## 3   The Model

### *3.1   Overview*

Agents are modeled as finite state automata (FSA) with a maximum number of feasible states. As with normal renditions of these automata, each agent has an initial state which is not contingent on the opponent they are playing, and each state describes both their action for that state, and their state transition contingent on the play of their opponent.

Agents begin with a uniform interaction environment (a null-graph) and within a period undergo at least some minimum number of interactions with other agents to play the IPD. Within each interaction, agents are able to influence the interaction environment by signaling to their opponent that they wish to break the interaction and reveal their positive or negative response to their opponent. If both players play positive signals, an edge is assigned between them, and the two agents will meet each other with higher probability in the future. The exact value of this probability is contingent on how many other agents each has already formed a link with.

---

[12][As in (Miller, 1996, p.91)] A Moore machine is defined by the four-tuple $\{Q, q_o, \lambda, \delta\}$ where $Q$ is the set of *internal states*; $q_o$ is the initial state; $\lambda$ is a mapping from each state to the subsequent action to be played $\lambda : Q \leftrightarrow S_i$, for example, in the PD, $S_i \in \{C, D\}$; and $\delta$ is the *transition function* that maps from the current internal state of the machine to the new internal state, contingent on the *opponent's* reported move, $\delta : Q \times S_{\sim i} \leftrightarrow Q$, $S_{\sim i} \in \{C, D\}$ being the opponent's reported move last period (in this case, for the PD).

In this way, the concept of 'partner-scarcity' is incorporated: though link-formation increases the probability that two agents will meet again, it does not guarantee it. Consequently, successful agents must either protect themselves completely from exploitative players through link formation, or display a depth of complexity in their strategy that can manage playing against undesirable opponents (or a combination of the two).

At the end of a period, total payoffs are determined for each agent, and an 'elite' fraction of the population is retained for the next period, with the remainder being replaced by new agents.[13] New agents are generated from a combination of existing elite behaviours and new behaviours (a type of learning) followed by mistake-making/innovation. Elite agents retain their links between periods (so long as they are to fellow elites) whereas entrants begin with no links, befitting the concepts of incumbency and network dynamism.

In this way, links are established within a period by mutual agreement between two agents. However, links can only be broken when an agent leaves the population after selection, severing all pre-existing links.

### 3.2   Details

Let $\mathbf{N} = \{1, \ldots, n\}$ be a constant population of agents and denote by $i$ and $j$ two representative members of the population. Initially, members of the population are uniformly paired to play the modified IPD game $\mathcal{G}$ described below. When two agents are paired together, they are said to have an *interaction*. Within an interaction, agents play the IPD for up to a maximum of $\tau$ iterations, receiving a payoff equal to the sum of the individual payoffs they receive in each iteration of the IPD. An interaction ends prematurely if *either* player plays a 'signal' thus unilaterally stopping the interaction. A strategy for a player $s$ describes a complete plan of action for their play within an interaction, to be explained presently. In addition to the normal moves of cooperate ($C$) and defect ($D$), an agent can also play one of two signal actions, $\#_s$ and $\#_w$ respectively. Thus, in any one iteration of the IPD, the action-set for an agents is $\{C, D, \#_s, \#_w\}$. As mentioned above, the playing of a signal by either player leads to the interaction stopping, possibly prior to $\tau$ iterations being reached. The playing of a signal can thus serve as an exit move for a player.

The interpretation of the two types of signal is as follows. Although initial pairing probabilities between all players are uniform random, agents can influence these interaction probabilities through the use of the signals.

---

Formally, let some agent $i$ maintain a preference vector,

$$\{f^i : f^i_j \in \{p_s, p_0, p_w\} \ \forall \ j \in \mathbf{N}/i\} \tag{2}$$

where $f^i_j$ is the preference status of agent $i$ towards agent $j$ and $p_s > p_0 > p_w$ are natural and denote *strengthen*, *untried* and *weaken* preferences respectively. Initially all entries are set to $p_0$ for all $j \in \mathbf{N}/\{i\}$. A probability vector $r_i$ for each agent is constructed from the preference vector by simple normalisation onto the real line,

$$\left\{r^i : r^i_j = \frac{f^i_j}{\sum f^i} \quad \forall \ j \in \mathbf{N}/\{i\}\right\} , \tag{3}$$

such that each opponent occupies a finite, not-zero length on the line $[0,1]$ with arbitrary ordering. Since we study here a model of mutual network/trust formation, preferences can be strengthened only by *mutual* agreement. Specifically, if agents $i$ and $j$ are paired to play the IPD, then when the interaction ends in iteration $t \leq \tau$,

$$f^i_j = f^j_i = \begin{cases} p_s & \text{if } s^i_t = s^j_t = \#_s , \\ p_w & \text{else} , \end{cases} \tag{4}$$

where $s^i_t$ denotes the play of agent $i$ in iteration $t$. That is, in all cases other than mutual coordinated agreement, the two agents will lower their relative likelihood of being paired again (though the playing of $\#_w$ might cause the interaction to end prematurely with the same result). Payoffs for each iteration of the PD are given by (5) below.

$$\mathcal{G}(\mathbf{I}, \mathbf{II}) = \quad \mathbf{I} \quad \begin{array}{c|cccc} & \multicolumn{4}{c}{\mathbf{II}} \\ & \#_s & C & D & \#_w \\ \hline \#_s & (0,0) & \multicolumn{2}{c}{\cdots} & (0,0) \\ C & & (3,3) & (0,5) & \\ D & \vdots & (5,0) & (1,1) & \vdots \\ \#_w & (0,0) & \multicolumn{2}{c}{\cdots} & (0,0) \end{array} \tag{5}$$

The playing of signals, is costly: the instantaneous cost for that period is the foregone payoff from a successful iteration of the IPD.

### 3.3   Example

Let two agents $i$ and $j$ be chosen to play (5) in some period and let maximum interaction length $\tau = 3$. Consider the following interaction,

| Iteration | $P_i$ | $P_j$ | $\pi_i$ | $\pi_j$ |
|-----------|-------|-------|---------|---------|
| 1 | $C$ | $C$ | 3 | 3 |
| 2 | $D$ | $C$ | 5 | 0 |
| 3 | $\#(s)$ | $C$ | 0 | 0 |
| $\sum \pi_x$ | | | 8 | 3 |

note that $i$ played an unrequited strengthen signal in the third (last) iteration; both players' interaction preference entries would be set to $p_w$.

### *3.4  Game Play*

In a *period* each agent is addressed once in uniformly random order to undergo $m$ interactions with players drawn from the rest of the population ($\mathbf{N}/\{i\}$). An agent is paired randomly in accordance with their interaction probability vector $r^i$ with replacement after each interaction. Preference and probability vectors are updated after every interaction.

Thus, it is possible that, having previously interacted with all agents, an agent retains only one preferred agent, whilst all others are non-preferred, causing a high proportion (if not all) of their $m$ interactions to be conducted with their preferred partner. However, it is to be noted that the value of $m$ is only a *minimum* number of interactions for an agent in one period, since they will be on the 'receiving end' of other agents' interactions in the same period. In this way, agents who incur an immediate cost of tie strengthening (foregoing iteration payoffs) can gain a long-term benefit through further preferential interactions.

At the end of $T$ periods, the population undergoes selection. A fraction $\theta$ of the population is retained (the 'elites'), whilst the remainder $(1-\theta)$ are replaced by new agents as described below. Selection is based on a ranking by total agent payoffs over the whole period. Where two agents have the same total payoff in a period, the older player remains.[14]

### *3.5  Agent Modeling*

Each agent is modeled as an $k$ (maximum) state FSA.[15] Since the interaction will stop immediately after either player plays the signal # each state must include two transition responses only: $R(C)$ and $R(D)$. For example, an agent's first three states might take the form (schematic representation given in Fig. 2) ,

| State | $P$ | $R(C)$ | $R(D)$ |
|-------|-----|--------|--------|
| 1 | C | 1 | 3 |
| 2 | C | 2 | 1 |
| 3 | D | 1 | 3 |

where the first state could be read as,

> '*play C next, if the opponent plays D, go to state 3; else, stay in state 1.*'

---

[14]Following SSA Smucker et al. (1994).

[15]To facilitate the computational modeling of this environment, agent strategies were encoded into binary format. See Miller et al. (2002) for an analogous description of this method for FSA.

An agent will have $k$ such states as part of their 'strategy'. By convention, the first state is taken as the initial one.

It can be shown[16] that the maximum number of states $\tau$ possible for an FSA playing some game with count $|R|$ feasible transition responses lasting at most $\tau$ rounds is simply,

$$k(R, \tau) = \sum_{t=0}^{\tau-1} |R|^t \ ,$$
(6)

hence, under this regime, the maximum interaction length $\tau$ and the number of opponent plays requiring a response defines the maximal FSA length. However, it is to be noted that there is no guarantee that all $k$ states for a given agent will be accessed (consider the example given immediately above, state 2 is present but is strategically redundant). In this way, FSA give a tangible sense of 'strategic complexity' when it comes to individual strategies. An agent who uses all of their $k > 1$ states as part of their strategy will no-doubt display a deeper strategy in action, than an agent who is playing merely (say) ALL-C. Of course, such complexity of strategy may or may not correspond to relative success in the population.[17]

After each period, a fraction $\theta$ will stay in the population, with the remaining agents being filled by new entrants. Here, the process of imitation and innovation/mistake-making is implemented via two foundational processes from the *genetic algorithm* (GA) literature.[18] Initially, two agents are randomly selected (with replacement) from the elite population. A one-point crossover operator is applied to each agent, and two new agents are formed. The strategy encoding (bit-strings) of these new agents then undergo point mutations at a pre-determined rate (5 bits per 1000). This process (random selection, crossover and mutation) continues until all the remaining spots are filled.

## 4 RESULTS & DISCUSSION

### 4.1 Uniform interactions

To begin, we study a static uniform interaction space to check any unwanted outcomes due to the modified IPD set-up. In this situation, rather than agents upgrading their preference vector after each interaction, the preference vector is uniform and unchanged throughout the model. In this way, the effect of the modification to the standard IPD framework can be analysed. Under such a scenario, the action set for each agent reduces to $\{C, D, \#\}$ since the signal action $\#$ has no interaction space interpretation, but still provides a means of prematurely ending the interaction (thus we may drop the sub-script).

---

[16]Consider a single-rooted logic tree where each node is a state, and each branch some transition. Without any re-use of states, the tree can be at most $\tau - 1$ nodes deep. Now observe that after the initial node (call this layer $t = 0$) each new layer will produce $r^t$ new nodes. The result follows.

[17]For example, it has been shown countless times before that the humble TIT-FOR-TAT (play $C$ until the opponent plays $D$, then switch to $D$ until the opponent plays $C$, then switch back to $C$, and so on.) strategy (and its variants) is often an extremely effective one against all manner of opponents, even though it can be represented by just two states!

[18]See, for example, Goldberg (2002) or for a non-technical introduction, Holland (1992).

To keep matters simple, we consider a model in which the maximum interaction length $\tau = 2$, which by (6) yields a maximum FSA state count of $k = 3$. Under these conditions, a strategy will be composed simply of a first play, and response plays to $C$ and $D$.

With uniform interaction probabilities, this model can be thought of as a modified evolutionary game theoretic framework. The probability of interacting with a certain agent type is directly equal to the proportional representation in the population of that type. Modification of the standard framework is due to the genetic algorithm approach, the crossover operator providing for imitation in addition to the more standard 'random' mutation operator. However, in terms of evolutionary stable strategies, this modification is insignificant.

In this setting, no evolutionary stable strategy will include $\#$ as a first play, since the payoff for such a strategy with any other agent is $0$.[19] This leaves strategies in the form of a triplet,

$$s : \{P_1, R(C), R(D)\} \ ,$$

where $P_1 \in \{C, D\}$ and $R(.)$ indicate subsequent plays in response to either $C$ or $D$ plays by the opponent $R(.) \in \{C, D, \#\}$. In all, 18 unique strategies can be constructed.

It is instructive to consider whether cooperative strategies might be evolutionary stable in this scenario. Clearly, a strategy $s_C : \{C, C, C\}$ will yield strictly worse payoffs than the strategy $s_D : \{D, D, D\}$ in a mixed environment of the two. It can be shown[20] that the strategies $s_{C\#} : \{C, C, \#\}$ and $s_{CD} : \{C, C, D\}$ (Figs. 3 and 4 , constitute the only two ESSs in an environment of $s_{D,D,D}$ only. However, both $s_{C\#}$ and $s_{CD}$ are themselves susceptible to attack by a 'mimic' agent such as $s_M = \{C, D, D\}$, which itself will yield to the familiar $s_D$. In this way, even with the added facility of being able to end the interaction prematurely, the only evolutionary stable strategy with respect to the full strategy space is that of $s_{D,D,D}$. Any intermediate resting place for the community will soon falter and move to this end (Fig. 5).

---

[19] The interaction would end after the first iteration, and $\mathcal{G}(\#_x|y) = 0$ for all $x \in \{s, w\}$ and $y \in \{C, D, \#(s), \#(w)\}$.

[20] See Lemma 1 in Appendix.

### *4.2   Uniform Interactions: Computational Results*

Computational experiments were run under uniform mixing as described above as a method of model validation. Given the maximum interaction length of $\tau = 2$, results for strategies present as a fraction of the total population are given in Figure 7 .[21] As predicted above, the model shows the clear dominance of $s_D$ under uniform mixing. Additionally, as predicted, the initial 'shake-out' periods ($t < 30$) gave rise to interesting wave-like strategic jostling. Agents playing cooperation first, and replying to $D$ with # were the first to have an early peak, if short-lived, which is not unexpected, since playing the signal is not the best-response to any subsequent play. Thereafter TFT-nice ($C$) peaked, but were soon overcome by the turn-coat type (who dominates TFT-nice). However, as the stock of $C$ players diminish, 'turn-coat' too, yields to the $D$-resp type strategies (such as TFT-nasty ($D$)).

We may conclude then, that the presence of the signal play (#) does little to affect *strategic* outcomes in the standard IPD set-up; defection still reigns supreme in the uniform IPD environment.

⇐ **Fig 7**
about here### *4.3   Non-uniform Interactions: Network Formation*

From the preceding analysis, a natural question arises: under what circumstances, if any, do networks emerge? Whilst it is possible to think of the network formation process as an aid to certain dominated behaviour (e.g. mutual cooperation), it is not obvious to what degree the network must shield certain players from hostile behaviour before their interaction community can be self-supporting.

Two factors that will clearly affect the propensity for cooperative networks to arise, are:

1. The 'impact' of the network on the interaction space; to what extent a strengthening signal actually changes mixing probabilities – formally, the values of the preference set $\{p_0, p_s, p_w\}$; and

2. The minimum number of interactions *within* a period over which two agents can exploit a beneficial relationship – formally, the value of $m$.

These two factors are related, since a low interaction impact may be compensated for by an high absolute number of interactions within a period.

The impact of network formation decisions by agents was parametrised in the computational experiments as follows,

$$p_w = (1 - \eta)^2 \quad \text{and} \tag{7}$$

$$p_s = (1 + \eta)^2 , \tag{8}$$

---

[21]Modelling notes: each plot-line represents the average result from 20 modelling runs where in each run: $n = 100$ and $m = 20$ with 20 agents replaced after each period, under roulette-wheel type selection of elites (with replacement) and $p(mutation) = 0.5\%$. Results were unchanged in substance for $m \in \{2, 5, 10\}$.

where $\eta \in [0, 1)$. The choice of the expression is somewhat arbitrary, however, the current specification retains symmetry about $p_0 = 1$ for all values of $\eta$ and by taking the squared deviation from 1, the ratio $p_s/p_w$ could be easily varied over a wide range. For example, by choosing $\eta = 0.8$, we yield $p_w = 0.04$ and $p_s = 3.24$, which gives a ratio of preferences in $f^i$ for agent $i$ for two agents with such values of 81. That is, the preferred agent (where an edge is assigned for the purposes of visualisation) will be met around 80 times more regularly than the less preferred agent when $i$ is being addressed.

To determine what conditions are favourable for network formation, a second computational experiment was conducted, this time 'turning up' the *interaction space impact* of any signalling play by the agents. Specifically, the network tuning parameter $\eta$ was varied in the range $[0.2, 0.95]$ together with the minimum interaction parameter $m$ over $[2, 20]$.

It was found that necessary conditions for sustainable network formation were $\eta \gtrsim 0.8$ and $m \gtrsim 10$. In terms of the population, these accord with a ratio of $p_s$ to $p_w$ (by (7) and (8)) of around 80 times,[22] and a minimum fraction of interactions per period of around 10% of the population.

**Table 1.** Mean values (over 20 trials) for final period (50) agent degree, $\langle d \rangle$ and fraction of PD plays where mutual cooperation resulted, $f(C, C)$. (see also Fig. 6).

| $m \setminus \eta$ | $\langle d \rangle$ | | | $f(C, C)$ | | |
|---|---|---|---|---|---|---|
| | 0.80 | 0.90 | 0.95 | 0.80 | 0.90 | 0.95 |
| 10 | 0.000 | 0.000 | 0.000 | 0.0000 | 0.0000 | 0.0000 |
| 14 | 0.004 | 0.001 | 0.391 | 0.0001 | 0.0000 | 0.0058 |
| 18 | 2.441 | 11.859 | 8.587 | 0.0286 | 0.1109 | 0.0735 |
| 20 | 7.959 | 11.073 | 9.548 | 0.0914 | 0.0941 | 0.1185 |

Further, the fraction of mutual cooperative plays (of all PD plays) moved in an highly correlated way with degree (see Table 1 and Fig. 6 ). It would appear, therefore, that network formation in this model is due to agents who play $C$ first, and $P[R(C)] = \#_s$.[23] A closer look at the dynamics of prevalent strategies under network forming conditions confirms this conclusion (see Fig. 8).

$\Leftarrow$ **Fig 6** about here

We study here an example $(m, \eta)$ combination at $m = 20$, and $\eta = 0.8$ (see Fig. 8 ). Four agent types are of interest (along with the summed $D$-responder types): the cooperative network forming type (`C-NET`); the defection network forming type (`D-NET`); and two types which engage in an highly asymmetric relationship – the opportunist (`D-OPP`) and so-called 'sucker' (`C-SCK`) types. Again, the periodic rise and fall of strategy types

$\Leftarrow$ **Fig 8** about here

---

[22]That is, an agent is 80 times more likely to interact with a preferred agent rather than a disliked agent in a given period.

[23]Recall, agents are free to form networks with any kind of behavioural basis.

is evident, but importantly, it can be seen that although `C-NET`, `D-NET` and `D-OPP` appear to co-exist for a time, it is only the cooperative network forming type who prevails in the long run.

To better understand these dynamics, a series of network snapshots for one representative network formation trial under the above conditions is shown in Figs. 9 and 10 . Here, at least four distinct phases are discernible.

**Phase 1: Amorphous connected (Figs. 9(a) and 9(b))**   The existence of many sucker types leads to a super network with high average degree. In this case, almost all of the cooperative types have formed links to at least one sucker type, whilst the opportunists are largely integrated into the super network, with a range of agent types as adjacent nodes.

**Phase 2: Segregated connected (Figs. 9(c) and 9(d))**   The network remains super connected, but clear segregation begins to occur, such that agent-to-agent edges become highly assortative. Fewer sucker types means that opportunists become competitive for activity in the network (e.g. Fig. 9(d)). Cooperative and defection communities subsequently establish themselves (higher intra-community connectivity).

**Phase 3: Segregated disjoint (Figs. 10(a), 10(b) and 10(b))**   The sucker type disappears, leading to a 'shake-out' in the population – the over-supply of opportunist types is rectified, with only those who were able to integrate with the defective community able to survive. The network is now dis-joint, with highly defined community characteristics. Further agent survival depends on raw mutual payoff characteristics.

**Phase 4: Homogeneous connected (Figs. 10(d) and 10(e))**   With the significantly higher intra-community payoffs yielded to the cooperative community, edges here become highly dense, approaching a complete component graph. The defective community disappears, with no possibility of infiltration into the cooperative community (see discussion below). New agents of cooperative network forming type are able to join and be integrated. Some sucker–opportunist relationships arise on margins but are short lived only.

### 4.3.1   Network Agent Types

A dissection of the prominent strategies that arose in the above experiment was conducted on period 13 (Fig. 9(c)) . A comparison of the network itself with the agent autopsies given in Fig. 11 makes clear the difference between each agent's activity in the network. Clearly, the interaction of the opportunist and sucker types (Fig. 11(b) and 11(c) respectively) will lead to tie-strengthening conditions, but with highly asymmetric payoff outcomes.

Importantly, however, the 'robust-C' type (agent 10 in period 13) is immune to this play by responding with $D$ to the opportunist's $D$ opening; a transition that works equally well for agent 10 when facing the robust-D type (agent 79 in period 13). For this reason, as can be seen in the agent networks presented so far, the cooperative types avoid tie-strengthening with either the robust-D or the opportunist types, which in both cases ensures adequate type-selection, but in the latter case, protects the cooperative network forming types from the opportunist shake-out that was inevitable with the decline of the sucker types in periods 13 to 17.

At the statistical level, these interactions are borne out in the periodic struggle of the initial network dynamics (see Fig. 8). The initial rise of the sucker types (establishing network ties to any other tie-strengthening agent) provides fertile ground for the opportunist types, who in turn, support the defection network types. However, over time, as each loses its respective 'feed-stock', network dynamics resolve in favour of the cooperative network forming types.

It is important to note that within this boundedly-rational framework, robust network formation is highly dependent on 'purity' of network structures. As can be seen in Fig. 12 , connected components that experience longevity must be able to attain more than the going 'outside' payoff rate of 2 per interaction.[24]

As can be identified, connected components that have a high proportion of sucker or opportunist types will yield large mean payoffs, but are very short-lived (rarely having mean agent ages greater than 5 periods) due to the volatile nature of payoff asymmetries. On the other hand, the cooperative networks who can overcome the short-term heterogeneous phase are very likely to retain higher than 2 average payoffs and so be positively selected for in the end-of-period strategy revision phase. Clearly, ensuring good 'discipline' within a cooperation network must be an high priority for the sustainability for the agents therein.

Interestingly, it appears from the data presented, that although predominantly defection type networks can yield very high payoffs, they will also suffer a type differentiation problem, mixing easily with the opportunist types. In the early stages of population dynamics, this a positive attribute since it will provide these types with high period payoffs through greater 'activity' (more plays of the IPD), ensuring their individual survival. However, over time, with the propensity for opportunist types to lose valuable payoff opportunities with sucker types, the defection networks yield strictly worse average payoffs than the 'outside' defection population, since they are necessarily sacrificing a unit of payoff every time they re-affirm/establish a link with a fellow defection network type.

---

[24]The payoff yield between two ALL-D types (for example) who play a two-iteration IPD game, gaining 1 in each iteration.

### 4.4 Multiple Equilibria & the Long Run

In the previous section, conditions were identified in which stable networks were formed under parsimonious agent specification ($\tau = 2$ implying $k = 3$) to enable correlation with established results in the analytic literature. Here, this constraint is relaxed and instead agents interactions of up to four iterations of the IPD game ($\tau = 4$) are considered and their long-run dynamics studied. Recall, by increasing the length of the IPD game, the maximal FSA state count increases markedly: for $\tau = \{3, 4\}$ maximum state count $k = \{7, 15\}$.

Previous conditions were retained, with $\eta = 0.8$ and $m = 20$, and each trial allowed to run for 1000 periods. Since a full description of the state is not feasible[25] we consider an aggregate description of two fundamental state characteristics, $f(C, C)$ – the fraction of plays in a period where mutual cooperation is observed (strategic behaviour); and $\langle d \rangle$ – mean agent degree (network formation). Results are presented for five long-run trials in Fig. 13. Under low interaction length the system moves within 100 steps to one of two stable equilibria – either a stable cooperation network is formed (as was studied in the previous section) or no network arises and a stable defection population sets in. However, as the interaction length increases (and so the associated complexity of behaviour that each agent can display), the dynamics become increasingly erratic, with multiple, apparently stable, equilibria visible in each case, but transient *transitions* between these equilibria observed. This situation is synonymous with that of *complex* system dynamics.

To better see this transition, the locations of the system in $f(C, C) - \langle d \rangle$ state-space were plotted (see Fig. 14). Here the transition from relatively well-defined attractors for $\tau = 2$ to complex dynamics at $\tau > 2$ is clear. Indeed, five stationary locations are visible in Fig. 5 with location I, II and V appearing to be transiently stable, with state trajectories both entering *and* leaving these locations, whilst locations III and IV appear to be absorbing for the system. Interestingly, these absorbing locations give rise to relatively similar average network formation, but different levels of cooperation, being low and moderate respectively.

Similarly, but with greater clarity, the dynamics of $\tau = 4$ shows very erratic behaviour (Fig. 5), appearing to have only two absorbing locations, IV and V, whilst each of I, II, and III appear to be transient. In this case, the absorbing locations are very different in character, being an almost complete graph, but similarly defection-based in the first case, or again, with high

---

[25]Consider that each time period, a population constitutes $n \times |s|$ bits, where $|s|$ is the length of a string needed to represent each agent's strategy, and the network $n(n-1)/2$ bits; taken together, gives rise to a possible $2^{n(n-1)/2+n|s|}$ states, which for $\tau = 2$ is $2^{9 \times 10^6}$! (It is possible to reduce this number by conducting automata autopsies, but the problem remains.)

participation, but markedly cooperative in the second.

Surprisingly, such complex dynamics arise in a relatively simple model of network formation. Recall, that the longest that any of the agent interactions can be in these studies was just two, three or four iterations of the modified Prisoner's Dilemma given in (5). To be very sure that such dynamics are not a consequence of the encoding of the automata themselves, an identical study was run with $\tau = 4$, but setting $\eta = 0$ such that all interactions would continue to be of uniform probabilities. However, in all cases, the system moved to a zero cooperation regime within the first 100 periods and remained there. Clearly then, we conclude that endogeneity of network formation is driving such complex dynamics as observed above.

### 4.5   Network Formation & Self-Organized Criticality

Next, given that the system displays complex dynamics for given values of $\tau$ and that network endogeneity is critical to such dynamics, it is natural to study the dynamics of network formation itself. For these purposes, the size (node count) of the principle (largest) network component that exists at the end of each period is studied. Example time-series for one $\tau = 4$ run are given in Fig. 15.   In the first figure, the size of the network itself is shown, whilst in the second, the first differences are given (i.e. $S_t - S_{t-1}$). It can be seen from this example, that changes in network size occur both on many time-scales and to various degrees. Such phenomena is synonymous with systems exhibiting critical behaviour (Bak et al., 1988); perturbations to the system cause mostly small, damped outcomes, but can occasionally have dramatic effects, likened to a 'domino-effect'.

To investigate this feature, frequency distributions of average network fluctuation sizes $D(\Delta S)$ were prepared for each interaction length. As can be seen in Fig. 16 the distributions appear to follow a power-law behaviour, that is of the form,

$$D(\Delta S) \sim \Delta S^{-\alpha} . \tag{9}$$

Such a relationship is often termed 'scale-free' since it indicates that the same overall systemic dynamics are operating on all spatial scales; small deviations build up over time and lead to large deviations in the long-run due to connectivity within the system.

Spatial self-similarity is one feature of critical systems, the second is that similar power-law scaling is observed in the temporal domain as well; normally manifesting as so-called '$1/f$' noise, which appears ubiquitous in nature.[26] A power spectrum was therefore prepared of the time-series network size to study this possibility.[27] Fig 17  gives the outcome of this analysis,

---

[26]Examples from the introduction to (Bak et al., 1988) include: light from quasars, the intensity of sunspots, the current through resistors, the flow of sand in an hour glass, the flow of the Nile river, and stock exchange price indexes.

[27]Suppose $s(t)$ is the (discrete) times-series of some network size data (as per Fig. 15(a)),

showing clear power-law scaling behaviour. Linear fits were prepared for the first 10 data points[28] with good agreement in all cases. Exponents of the relationship,

$$S(f) \sim f^{-\alpha} \; , \tag{10}$$

were found all found to be $-1.8 \pm 0.1$.

Taken together, the spatial and temporal fingerprints of criticality observed in the network formation dynamics, indicate that the system is indeed very capable of the kind of complex dynamics observed and discussed above, and that the network formation appears to be a key factor in such behaviour. Furthermore, as has been proposed by various authors, rather than such criticality arising from fine tuning of system parameters such as occurs in designed critical industrial systems (e.g. nuclear fission reactors), the system appears to naturally move towards this critical state, and keep returning to it over time. It is for this property that authors such as Bak et al. (1988) have termed such phenomena 'self-organized criticality'. Indeed, it appears that such phenomena is a strong indicator of complex dynamics, and may indeed be the necessary system state to give rise to the kind of non-equilibrium processes observed in various dissipative systems.[29]

The existence of such dynamics in economic systems has recently received growing interest.[30] Indeed, power-law behaviour on both a macro (Canning et al., 1998; Devezas and Modelski, 2003) and micro- interactions scale (Arenas et al., 2000; Scheinkman and Woodford, 1994) has been incorporated into both models and empirical evidence, and some assert is fundamental to our understanding and thus modelling of economic systems (Arthur, 1994). Hence, the existence of such dynamics and features in the present model is a pleasing indication that significant features of realistic network formation contexts has been incorporated.

## 5  Conclusions

In contrast to previous attempts at capturing the dynamics of strategic network formation, the present model provides a relatively simple foundation, but powerfully rich behavioural and topological environment within which to study the dynamics of strategic network formation. Moreover, in contrast to previous dynamic and strategic network models, by incorporat-

---

then using Matlab a Fast-Fourier-Transform, $F(s)$ was performed with $N = 2^8$ points, followed by the standard power-function, $FF'/N$, where $F'$ is the complex conjugate of $F$. Figures show the resultant power spectra without the first constant-shift term $f(0)$, and are cut below $S(f) \leq 5 \times 10^{-4}$.

[28] Fitting power-law models has received some interesting study in recent times due to difficulties in forming goodness-of-fit tests etc. Here we follow Goldstein et al. (2004) in form, fitting the linear specification to only a selection of the primary points, thus avoiding undue bias in the tails (which represent a very small mass of the spectrum).

[29] See for example, Langton (1992) for a discussion on this point.

[30] See for example, Krugman (1996).

ing the network formation decision-process into individual agent strategies, a rich ecology of agent types and consequent network topologies was observed. Significantly, this model suggests that the network formation process must deliver relatively symmetric payoffs to network members. If this is not true, networks formed will likely be heterogeneous in nature, with disruptive edge formation and breaking sequences which can effectively destroy any benefits that the network might have conferred on members (e.g. the opportunist-sucker network volatility of phases I and II mentioned above).

Analytical and subsequent computational components of the present paper indicate that in this simple modified IPD set-up, cooperation is not sustainable without the additional benefits conferred by the type-selection and type-protection network externalities. Specifically, agents require at least some level of repetition of interaction within the current population to gain sufficient incentives to form the network; and secondly, the 'impact' of the network on the interaction space was found to be a necessary condition for cooperative network formation, with network emergence only observed when the network allows for relatively high (though not complete) discrimination (probabilistically) from the wider population.

Furthermore, the dynamical properties of the present model have been investigated and indicate that even with parsimonious descriptions of boundedly-rational agent strategies, complex dynamics are observed, with multiple and transient stationary locations a feature of the state space. These dynamics increased in complexity with increasing interaction length. Additionally, evidence on the fluctuations in the size of networks over time indicates that the network formation and decay processes themselves are likely the main driving force behind the complex system dynamics, with both spatial and temporal scaling behaviour indicating the existence of so-called 'self-organized criticality'. To this author's knowledge, this is the first strategic network formation model to produce and study such complex dynamics. Such observations clearly raise tantalising avenues for future work; I shall raise a selection in finishing: do realistic cooperative networks display complex dynamics? if not, what mechanism of agency overcomes such instability? if networks can be shown to have such dynamics (admittedly these data are still largely out of reach) what are the implications for supporting cooperative institutions? and finally, given an autonomous, locally interacting world, how should the social planner intervene in such networks to pursue welfare maximizing aims?

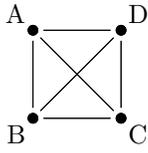

**Figure 1.** Uniform matching, $n = 4$.

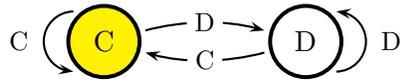

**Figure 2.** Example FSA: TIT-FOR-TAT.

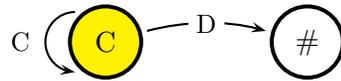

**Figure 3.** Agent strategy $s_{C\#} : \{C, C, \#\}$.

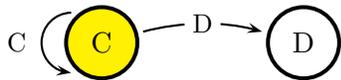

**Figure 4.** Agent strategy $s_{CD} : \{C, C, D\}$

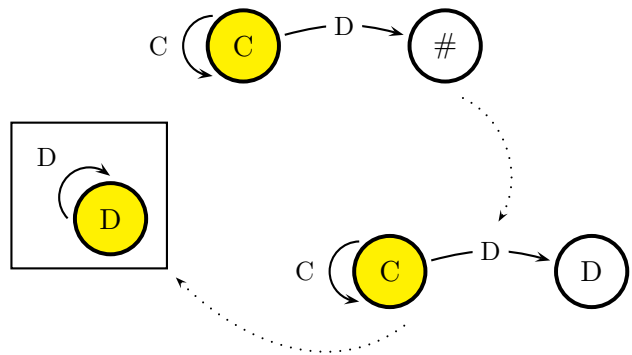

**Figure 5.** (Clock-wise from top) $s_{C\#}$ is stable in the presence of $s_D$ only but is susceptible to attack by $s_M$ which in turn will yield to $s_D$, the final evolutionary stable strategy component.

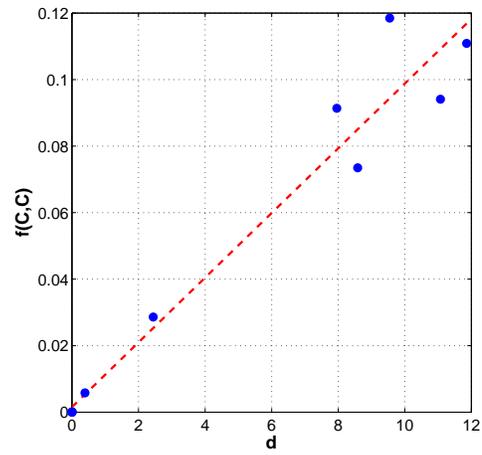

**Figure 6.** Fraction of mutual cooperation per PD vs. average agent degree under each condition listed in Table 1. Line (−−−) indicates simple OLS regression expected values.

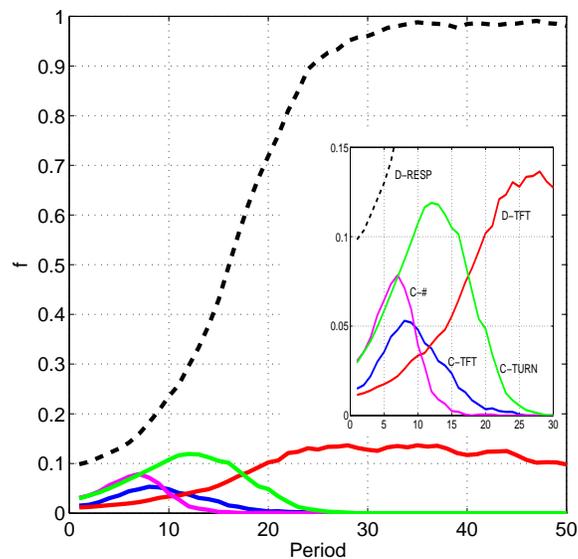

**Figure 7.** Population fraction of strategies under uniform mixing: (D-TFT) 'nasty' TIT-FOR-TAT; (C-TFT) 'nice' TFT; (C-#) plays $C$, returns $D$ with #; (C-TURN) 'turn-coat', plays $C$, but $D$ thereafter; (D-RESP) sum of all players who play $D$ and return $D$ with $D$.

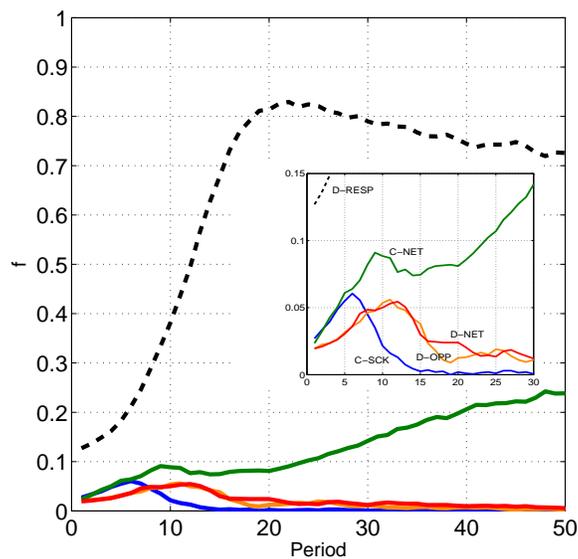

**Figure 8.** Example mean population fraction types playing various strategies (over 20 trials) under necessary network formation conditions ($m = 20, \eta = 0.8$): (C-NET) robust cooperation; (D-NET) robust defection; (D-OPP) opportunist; (C-SCK) 'sucker'; and (D-RESP) mutual defector types.

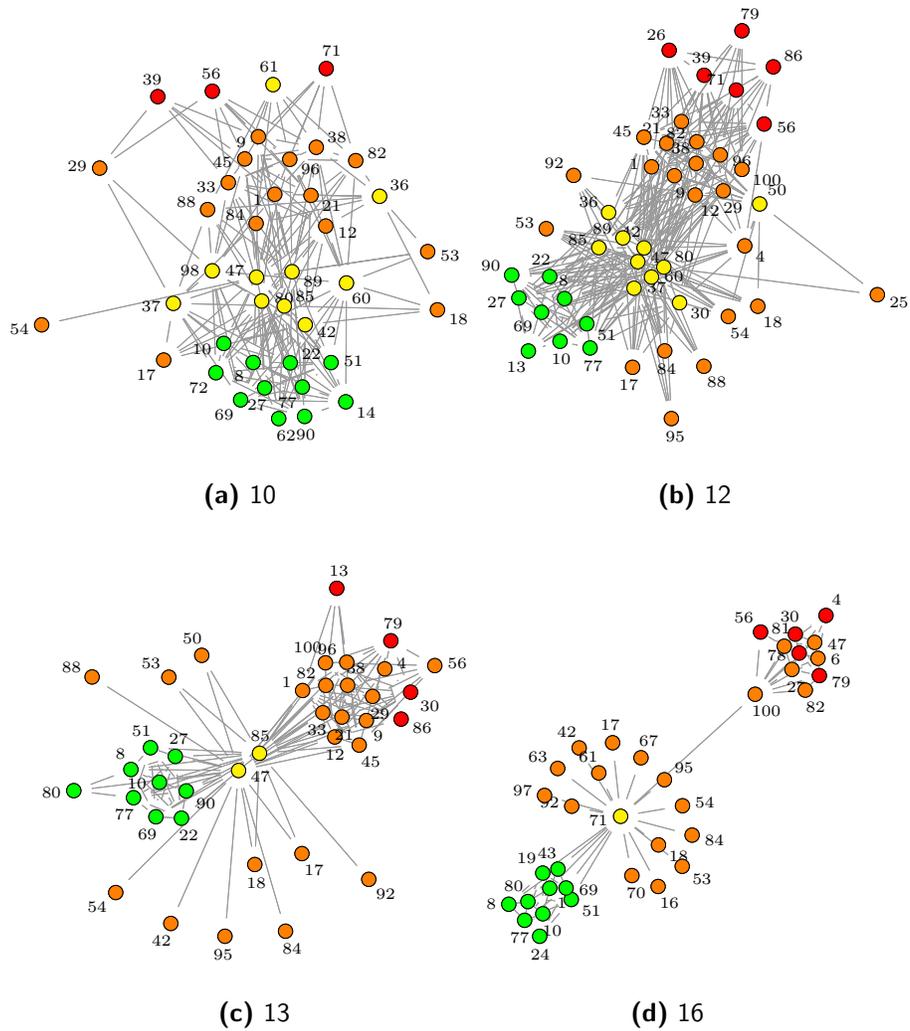

**(a)** 10

**(b)** 12

**(c)** 13

**(d)** 16

**Figure 9.** Example network dynamics ($m = 20, \eta = 0.8$): network state at end of indicated period; agent ID show next to each node; agent coloring as follows – (●) robust cooperative; (●) robust defection; (●) opportunist; and (●) 'sucker' (see text for explanation).

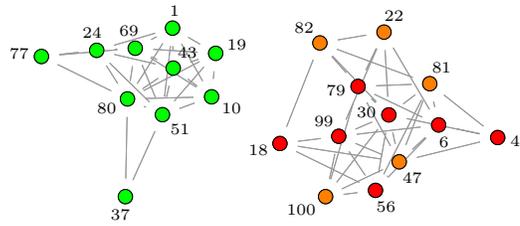

**(a)** 17

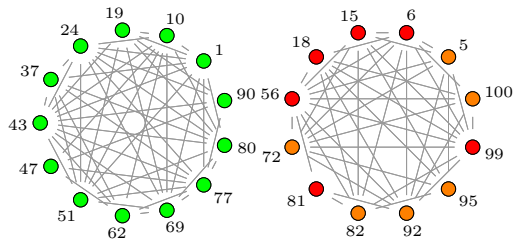

**(b)** 18

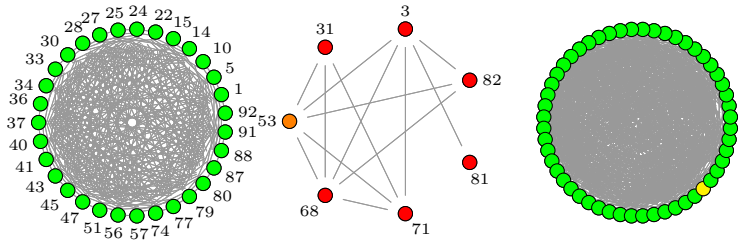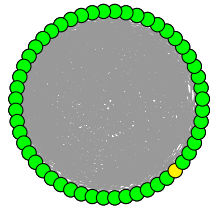

**(c)** 28      **(d)** 35

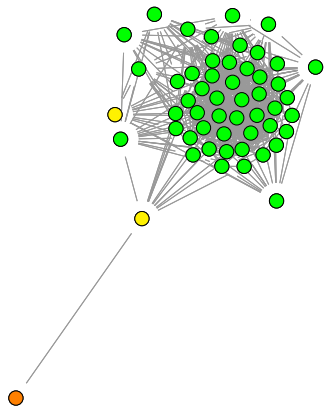

**(e)** 40

**Figure 10.** Example network dynamics (cont.): network state at end of indicated period (see Fig. 9 for explanation). *NB: Agent ID omitted for periods 35 and 40.*

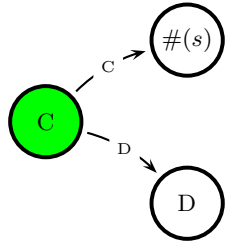

**(a)** ID 10 'robust-C'

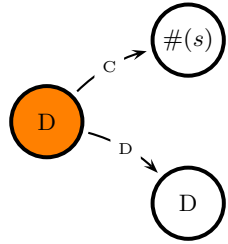

**(b)** ID 92 'opportunist'

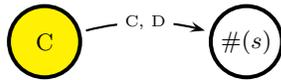

**(c)** ID 85 'sucker'

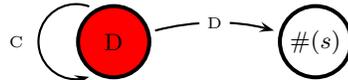

**(d)** ID 79 'robust-D'

**Figure 11.** Example types visible in the network of period 13 above (see Fig. 9).

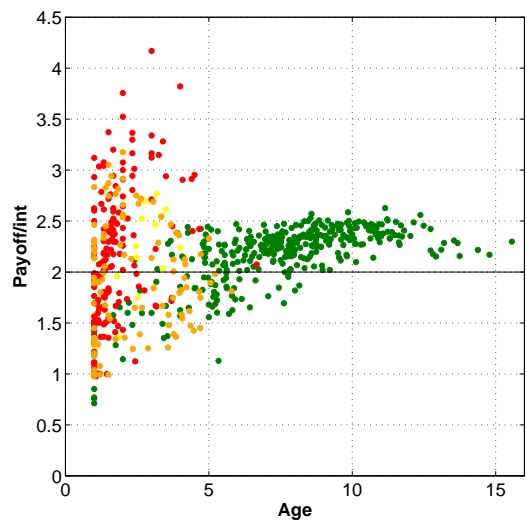

**Figure 12.** Scatter plot, mean Payoff per interaction vs. mean Age for each point which represents a single connected component. Coloring indicates dominant ($> 50\%$) type in each component: (●) 'suckers'; (●) 'opportunists'; (●) defection network builders; and (●) cooperation network builders. NB: line at interaction payoff 2 indicates expected payoff for non-connected $D$-types. (Other parameters as for Fig. 8).

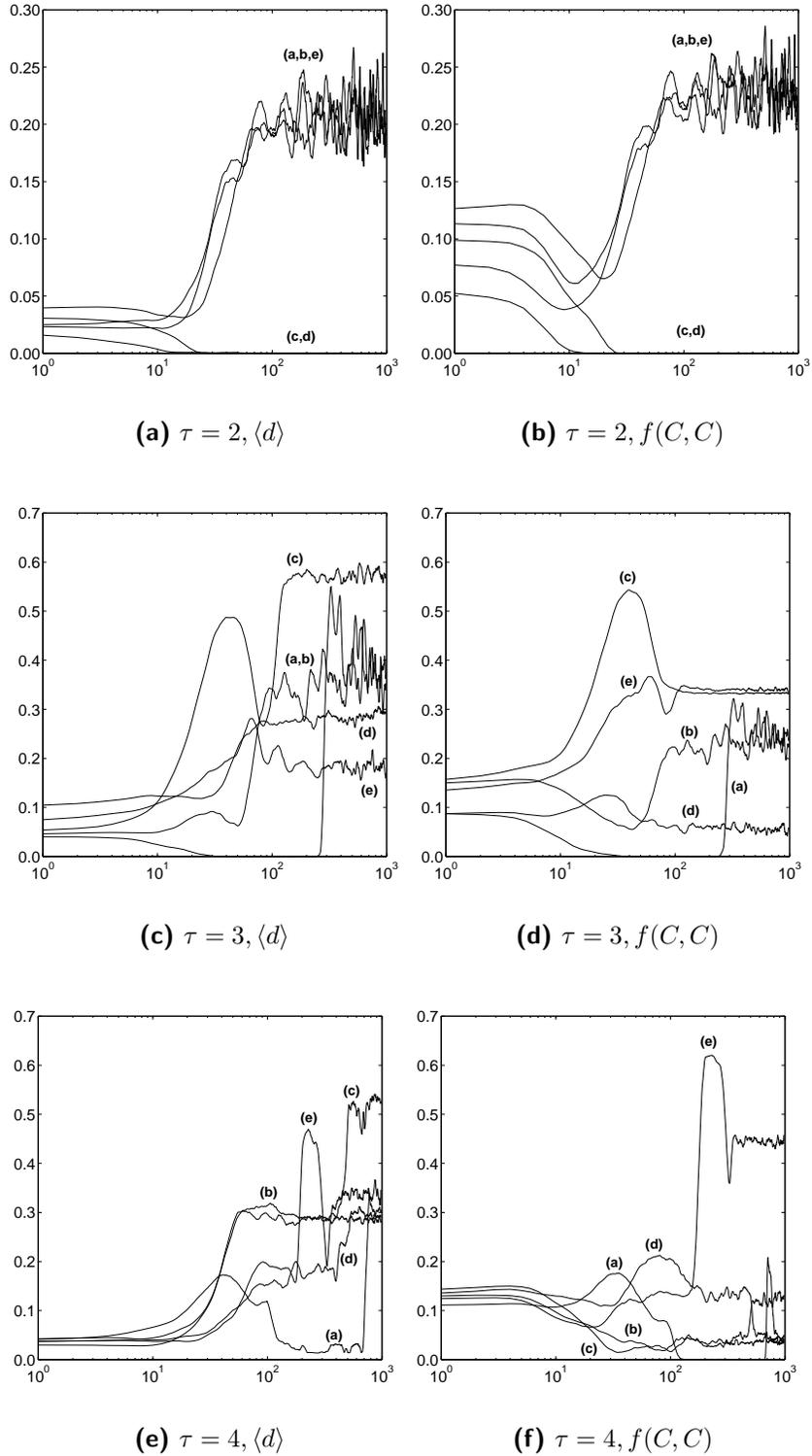

**Figure 13.** Long-run system dynamics under different maximum interaction lengths indicating increasing complexity; five trials shown at each value of $\tau$ (data smoothed over 20 steps).

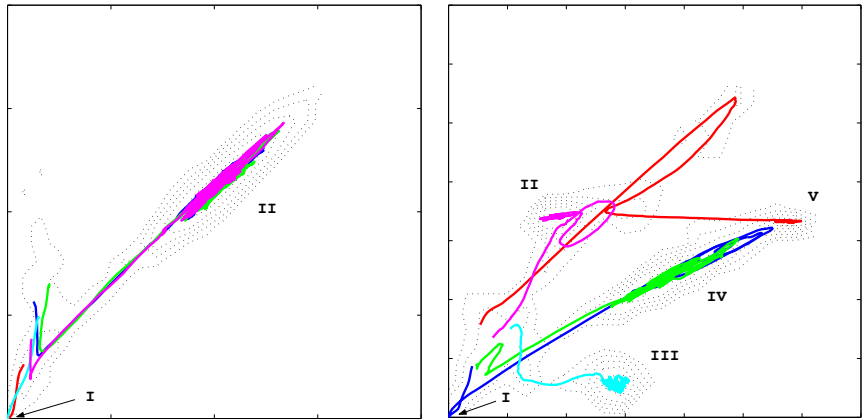

**(a)** $\tau = 2$       **(b)** $\tau = 3$

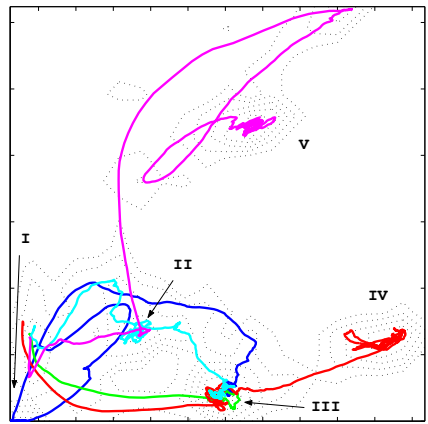

**(c)** $\tau = 4$

**Figure 14.** State transitions, $f(C, C) - \langle d \rangle$ space. Contours indicate ($\log_{10}$) density of time periods spent at $(f, \langle d \rangle)$ points over all trials; lines represent smoothed trial trajectories (20 steps), coloring shows each trial.

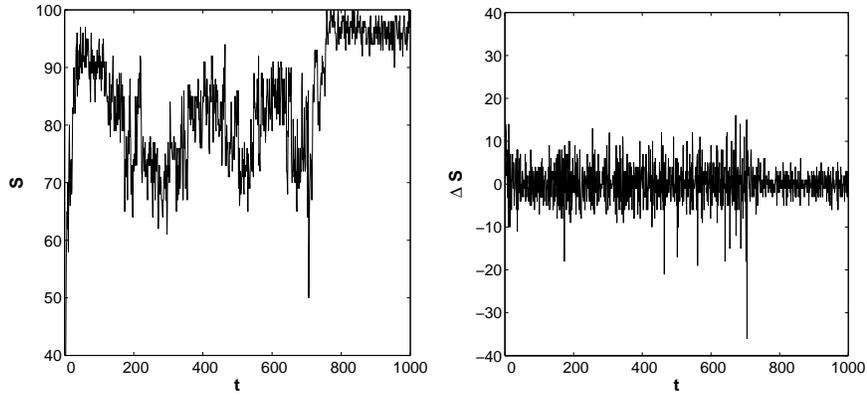

**Figure 15.** Example time series of principle component size (a) and change in size (b) for $\tau = 4$.

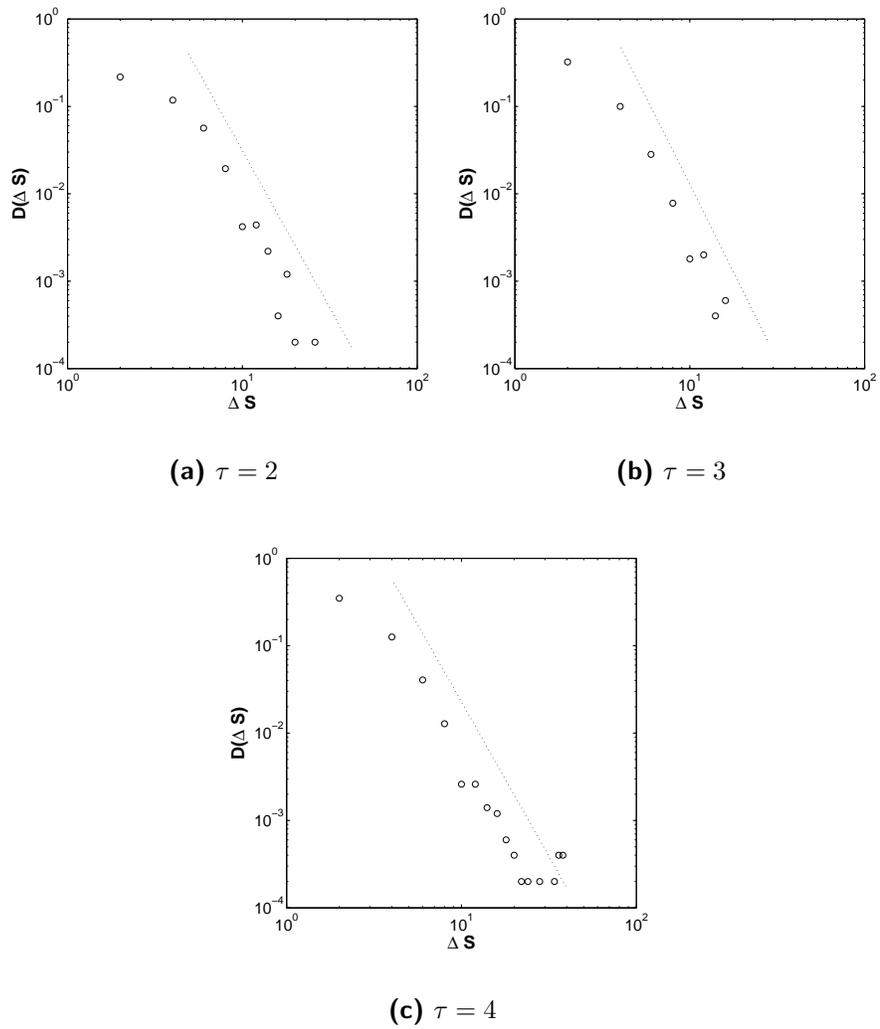

**(a)** $\tau = 2$        **(b)** $\tau = 3$

**(c)** $\tau = 4$

**Figure 16.** Mean principle component network size distribution for each interaction length. Lines given as guide only.

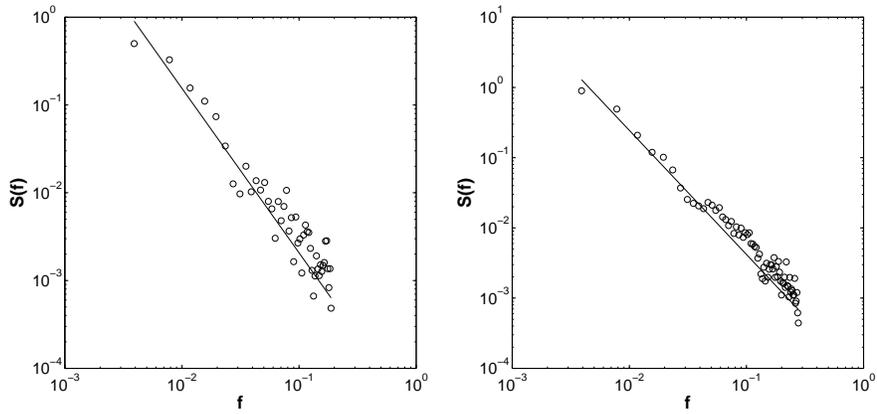

**(a)** $\tau = 2$        **(b)** $\tau = 3$

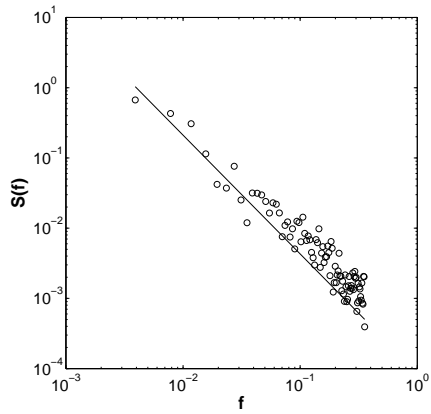

**(c)** $\tau = 4$

**Figure 17.** Mean power spectra of principle component network size for each interaction length. Lines represent power-law fits to first 10 points with $\alpha = -1.9, -1.8, -1.7$ for $\tau = 2, 3, 4$ respectively.

Appendix 1

Lemma 1 *For a population playing the IPD as given in $\mathcal{G}$ under uniform interaction probabilities and maximum FSA state length $\tau = 2$, the strategy triplet $s_D : \{D, D, D\}$ is the only evolutionary stable strategy (ESS).*

*Proof*  To begin with, consider a population consisting of only two types of agents, where one type is $s_D$ and the other the strategy triplet $s_S : \{P_1^S, R^S(C), R^S(D)\}$ where $P_1^S$ is the agent's play in state one and $R^S(x)$ represents the agent's play in response to opponent's play $x$ in the preceeding iteration. For convenience, call these types **D** and **S** respectively.

Define the total payoff to an agent $X$ undergoing an interaction with agent $Y$ to be $\Pi(X|Y)$ and note that,

$$\Pi(\mathbf{D}|\mathbf{S}) = \mathcal{G}(D|P_1^S) + \mathcal{G}(D|R^S(D)) \tag{11}$$
$$\Pi(\mathbf{D}|\mathbf{D}) = 2 \times \mathcal{G}(D|D) \tag{12}$$
$$\Pi(\mathbf{S}|\mathbf{D}) = \mathcal{G}(P_1^S|D) + \mathcal{G}(R^S(D)|D) \tag{13}$$
$$\Pi(\mathbf{S}|\mathbf{S}) = \mathcal{G}(P_1^S|P_1^S) + \mathcal{G}(R^S(P_1^S)|R^S(P_1^S)) \tag{14}$$

Further, let $\alpha$ be the proportion of type **S** in the population. Then, the expected interaction payoffs for each agent type with uniform mixing is given by,

$$E[\Pi(\mathbf{S})] = \alpha\Pi(\mathbf{S}|\mathbf{S}) + (1-\alpha)\Pi(\mathbf{S}|\mathbf{D}) \tag{15}$$
$$E[\Pi(\mathbf{D})] = (1-\alpha)\Pi(\mathbf{D}|\mathbf{D}) + \alpha\Pi(\mathbf{D}|\mathbf{S}) . \tag{16}$$

Now suppose $\alpha \to 1$, if **S** is to be stable in the presence of **D** then it follows from (15) and (16) that,

$$\Pi(\mathbf{S}|\mathbf{S}) \geq \Pi(\mathbf{D}|\mathbf{S}), \tag{17}$$

which by (11) to (14) becomes,

$$\mathcal{G}(P_1^S|P_1^S) + \mathcal{G}(R^S(P_1^S)|R^S(P_1^S)) \geq \mathcal{G}(D|P_1^S) + \mathcal{G}(D|R^S(D)) . \tag{18}$$

Now suppose that $P_1^S = C$ then by payoffs given in (5), (18) becomes,

$$\mathcal{G}(R^S(C)|R^S(C)) \geq \mathcal{G}(D|R^S(D)) + 2 , \tag{19}$$

which implies that $R^S(C)$ must be $C$ and that $R^S(D) \neq C$. Hence, there are only two candidates for **S**, namely $s_{CD} : \{C, C, D\}$ and $s_{C\#} : \{C, C, \#\}$. However, both $s_{CD}$ and $s_{C\#}$ are not stable in the presence of the mimic agent $s_M : \{C, D, D\}$ which itself does not satisfy the condition given in (18).

Suppose on the other hand that $P_1^S = D$. This would imply (by substitution of payoffs into (18)) that,

$$\mathcal{G}(R^S(D)|R^S(D)) \geq \mathcal{G}(D|R^S(D))$$

which has no solution unless $R^S(D) = \{D, \#\}$.

Now, consider a population where $\alpha \to 0$ and suppose there exists some strategy $\mathbf{S}'$ such that $E[\Pi(\mathbf{S}')] \geq E[\Pi(\mathbf{D})]$ which again yields,

$$\mathcal{G}(P_1^S|D) + \mathcal{G}(R^S(D)|D) \geq 2 \ . \tag{20}$$

If $P_1^S = C$ (20) becomes,

$$\mathcal{G}(R^S(D)|D) \geq 2$$

which has no solution. Likewise, if $P_1^S = D$ we have,

$$\mathcal{G}(R^S(D)|D) \geq 1$$

which has a solution only if $R^S(D) = D$, completing the proof.   $\square$